\documentclass[aps,prb,twocolumn]{revtex4}
\usepackage{amssymb}
\usepackage{graphicx}

\begin{document}

\title{Nutation versus angular dependent NQR spectroscopy and the impact of
underdoping on charge inhomogeneities in YBa$_2$Cu$_3$O$_y$}
\author{Rinat Ofer and Amit Keren}
\affiliation{Physics Department, Technion-Israel Institute of Technology, Haifa 32000,
Israel}
\date{\today}

\begin{abstract}
We describe two different nuclear quadrupole resonance (NQR) based
techniques, designed to measure the local asymmetry of the internal electric
field gradient $\eta $, and the tilt angle $\alpha $ of the main NQR
principal axis $\textbf{\^{z}}$ from the crystallographic axis $\textbf{\^{c}}$. These techniques use the
dependence of the NQR signal on the duration of the radio frequency (rf)
pulse and on the direction of the rf field $H_{1}$ with respect to the
crystal axis. The techniques are applied to oriented powder of YBa$_{2}$Cu$%
_{3}$O$_{y}$ fully enriched with $^{63}$Cu. Measurements were
performed at different frequencies, corresponding to different in-plane
copper sites with respect to the dopant. Combining the results from both
techniques, we conclude that oxygen deficiency in the chain layer lead to a
rotation of the NQR main principal axis at the nearby Cu on the CuO$_{2}$
planes by $\alpha\simeq20^{\circ}\pm5^{\circ}$. This occurs with no change to $%
\eta $. The axis rotation associated with oxygen deficiency means that there
must be electric field inhomogeneities in the CuO$_{2}$ planes only in the
vicinity of the missing oxygen.
\end{abstract}

\maketitle

\section{Introduction}

Hole doping the CuO$_{2}$ planes of the cuprates creates natural
inhomogeneities. Determining the origin and form of these charge
inhomogeneities is essential for our understanding of their transport
properties. There are several theoretical works suggesting that phase
separation is a characteristic of the Hubbard model, and therefore an
intrinsic property of the CuO$_{2}$ planes.\cite{strips} Moreover, it has been
suggested that the phase separation, possibly in the form of strips, can
explain many of the unusual properties of the cuprates and leads to the
superconductivity (for example, Ref. \cite{strips2}).

Indeed, some of these materials show phase separation with segregated
hole-rich and hole-poor regions. Such inhomogeneities were found in La$%
_{2-x} $Sr$_{x}$Cu$_{1}$O$_{4}$ by several methods such as nuclear
quadrupole resonance (NQR),\cite{Singer} neutron scattering,\cite{Yamada}
and muon spin relaxation ($\mu $SR).\cite{MuSR1} In addition, Scanning tunnel microscope (STM)
experiments on underdoped Bi$_{2}$Sr$_{2}$CaCu$_{2}$O$_{y}$ (Bi-2212) showed
local density-of-states modulations\cite{McElroy2003} and inhomogeneity of
the superconducting gap on the samples' surface that can be associated with
the distribution of holes in the planes.\cite{STM1}

In YBa$_{2}$Cu$_{3}$O$_{y}$ (YBCO), phase separation was found at very low
doping levels, up to YBCO$_{6.35}$, with neutron scattering from phonons
related to charge inhomogeneity.\cite{Mook} $\mu $SR measurements show the
existence of a spin-glass phase for a similar doping range.\cite{Sanna} The
highest doping in which magnetic order was found in YBCO was in YBCO$_{6.6}$\cite{Mook2008} using neutron scattering. $^{89}$Y NMR study in YBCO$_{y}$
for $y=7$ and $6.6$ showed no phase separation at all.\cite{Bobroff}
Direct detection of charge inhomogeneity in YBCO$_{y}$ is nearly impossible
since STM measurements are very difficult due to oxygen loss in vacuum, and
surface problems. Various STM experiments in this compound showed an
inhomogeneous \cite{Edwards} or relatively homogenous surface,\cite{Yeh}
depending on the surface preparation procedure. In these circumstances NQR
might be the only tool to probe charge inhomogeneities in YBCO. Moreover,
NQR experiments are sensitive to the charge distribution in the bulk and not
just on the surface.

In this paper, we present NQR measurements on $^{63}$Cu enriched YBCO. This
YBCO compound has narrow NQR resonance lines, which allows us to distinguish
between different in-plane copper [Cu(2)] and associate each line with a
local environment. We then apply two additional NQR based methods called
nutation spectroscopy and angle dependent NQR (ADNQR). The two methods are
set to measure the quadrupole interaction asymmetry parameter $\eta $ and
the tilt angle $\alpha $ of the NQR main principle axis $\widehat{\mathbf{z}}
$ from the crystal axis $\widehat{\mathbf{c}}$. This allows us to measure
charge inhomogeneity in the CuO$_{2}$ plane. The main advantage of these
techniques is that they can be employed at every frequency in the NQR
spectrum. Therefore, unlike NMR, $\eta $ and $\alpha $ can be determined for
each ionic environment separately. Our main conclusion is that for Cu(2) in
unit cells with no oxygen deficiency $\alpha \simeq 0$ and $\eta \simeq 0.$
However, for every Cu(2) in a unit cell where an oxygen is missing, and
\emph{only} in these unit cells, $\alpha \simeq 20^{\circ }$ and $\eta
\simeq 0$. Such a tilt of $\widehat{\mathbf{z}}$ from the crystal axis $%
\widehat{\mathbf{c}}$ in special unit cells implies charge inhomogeneities
in the CuO$_{2}$ plane only next to oxygen deficiency.

The paper is organized as follows. In Sec.~\ref{NQR} we give a general
description of NQR and then introduce the less known experimental methods:
the nutation spectroscopy and the ADNQR. Theoretical calculations for these
techniques are also shown. In Sec.~\ref{Setup} we describe the
experimental setup. The experimental results are presented in Sec.~\ref%
{Results}. Finally the summary and conclusions on the experimental
techniques and their application to YBCO are presented in Sec.~\ref%
{Conclusions}.

\section{The experimental techniques}

\label{NQR}

\subsection{NQR}

In an NQR experiment one uses an rf magnetic pulse in order to cause
transitions between the nuclear energy levels. Nuclei with spin $I>1/2$ can
be viewed as positively charged oval objects. As a result, their energy
inside a solid depends on their orientation in the electrostatic potential $%
V(\mathbf{r})$ generated by the other nuclear and electronic charges. When
the nuclear poles are close to positive charges their energy is high, and
when the poles are close to negative charges the energy is low. The energy
difference between different orientations is determined by the electric
field gradient (EFG) tensor $V_{ij}=\frac{\partial ^{2}V}{\partial
x_{i}\partial x_{j}}$ at the position of the Cu nuclei. The directions can
be chosen so that $V_{ij}$ is diagonal. These directions are known as the
principal axis of the EFG. Due to Laplace equation ($V_{xx}+V_{yy}+V_{zz}=0$%
) the NQR Hamiltonian is determined by only two parameters, $\nu _{q}$ and $%
\eta $, and is given by:
\begin{equation}
\mathcal{H}_{Q}=\frac{\hbar \nu _{q}}{6}[3I_{z}^{2}-I^{2}+\eta
(I_{x}^{2}-I_{y}^{2})]  \label{HQ}
\end{equation}%
where $I_{\alpha }$ are the nucleus spin operators; $\nu _{q}$ is a
frequency scale, determined by the main EFG component $V_{zz}$, and the
quadruple moment of the nucleus $Q$;
\begin{equation}
\eta =\frac{V_{xx}-V_{yy}}{V_{zz}}  \label{etaDef}
\end{equation}%
is the asymmetry parameter with the convention $|V_{zz}|\geq |V_{yy}|\geq
|V_{xx}|$, and therefore $0\leq \eta \leq 1$.

For the spin 3/2 copper nucleus, the NQR Hamiltonian has two doubly
degenerate energy levels, and therefore only one resonance frequency, given
by:
\begin{equation}
f_{NQR}=\hbar \nu _{q}\sqrt{1+\frac{\eta ^{2}}{3}}.  \label{resonance}
\end{equation}%
Since $\nu _{q}$ is proportional to $V_{zz}$, it holds information on the
charges surrounding the nucleus. Therefore nuclei sitting in different
electronic environments will have different resonance frequencies. $\eta $
is a parameter that holds information on the anisotropy of the surrounding
charges in the $x$-$y$ plane. In the case of axial symmetry, $\eta =0$. It
is clear from Eq.~(\ref{resonance}) that in the standard NQR experiment
where only the resonance frequency is measured, $\eta $ and $\nu _{q}$
cannot be separately determined. However, the properties of the rf field
turn out to be handy.

The rf pulse Hamiltonian is
\begin{equation}
\mathcal{H}_{rf}=\gamma \hbar \mathbf{H}_{1}\cdot \mathbf{I}\sin \omega t,
\label{Hamrf}
\end{equation}%
where, in general, the rf field can be described as
\begin{equation}
\mathbf{H}_{1}=H_{1}\hat{\mathbf{r}}\qquad \hat{\mathbf{r}}=\left[ \sin
(\theta )\cos (\phi ),\sin (\theta )\sin (\phi ),\cos (\theta )\right] ,
\label{Hrf}
\end{equation}%
where $\theta $ and $\phi $ the polar angles relating the coil axis to the
quadrupolar reference frame.

The signal in the coil after a time $t$ from an on-resonance rf pulse ($%
\omega =\omega _{Q}\equiv 2\pi f_{NQR}$) with a duration $t_{p}$ is given by
\cite{pratt}
\begin{equation}
I(t_{p},t,\theta ,\phi ,\eta )\propto \lambda (\theta ,\phi ,\eta )\sin
[\lambda (\theta ,\phi ,\eta )\omega _{1}t_{p}]\sin (\omega _{Q}t)
\label{Ibefore}
\end{equation}%
where $\omega _{1}=\gamma H_{1}$ ($\gamma $ is the nucleus gyromagnetic
ratio), and $\lambda $ is an angular factor given by
\begin{equation}
\lambda (\theta ,\phi ,\eta )=\sqrt{%
r_{x}^{2}a_{x}^{2}+r_{y}^{2}a_{y}^{2}+r_{z}^{2}a_{z}^{2}},  \label{lambda}
\end{equation}%
where%
\begin{equation}
\mathbf{a}=\frac{1}{2\sqrt{3+\eta ^{2}}}(\eta +3,\eta -3,2\eta ).  \label{a}
\end{equation}%
Hence, the signal at time $t$ is a function of both the duration of the rf
pulse and the polar angles between the rf direction and the EFG. In the
following sections we review two methods that use these parameters to
determine $\eta $, and the angle $\alpha $ between $\widehat{\mathbf{z}}$ and
$\widehat{\mathbf{c}}$.

\subsection{Nutation spectroscopy}

\label{sectionnutationtheory}

The nutation spectroscopy NQR, developed by Harbison \emph{et al.}\cite{Harbison,Harbison2} is a method in which the signals are measured as
a function of the duration of the rf excitation pulse $t_{p}$. This method
is used to determine the asymmetry parameter $\eta $. Integrating Eq.~(\ref%
{Ibefore}) over the angles for isotropic powder and Fourier transforming
with respect to $t_{p}$ gives a powder pattern line shape that is described
by
\begin{eqnarray}
I(\omega _{p},t,\eta )\propto \int_{0}^{2\pi }d\phi \int_{0}^{\pi }\sin
\theta d\theta \lambda (\theta ,\phi ,\eta ) \nonumber
\end{eqnarray}
\begin{eqnarray}
\times\int_{-\infty }^{\infty }e^{i\omega _{p}t_{p}}dt_{p}\sin (\lambda (\theta
,\phi ,\eta )\omega _{1}t_{p})\sin (\omega _{Q}t).\label{I}
\end{eqnarray}
\begin{figure}[tbp]
\begin{center}
\includegraphics[width=8cm]{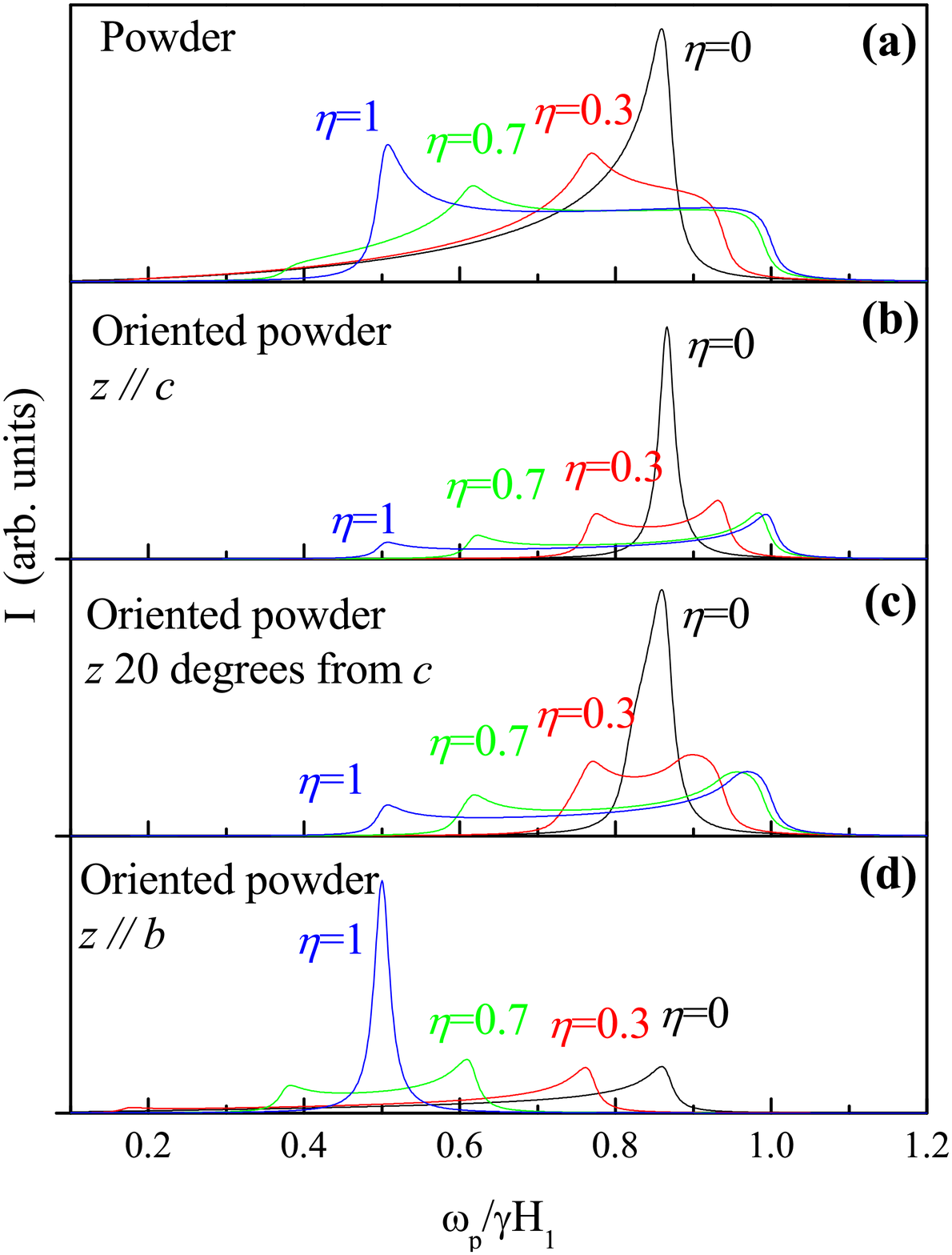}
\end{center}
\caption{(Color online) Calculated on-resonance nutation spectra for: (a)
powder samples. (b) Perfectly oriented powder where $\widehat{\mathbf{z}}%
\Vert \widehat{\mathbf{c}}$. (c) Oriented powder where $\widehat{\mathbf{z}}$
has up to $20^\circ$ angle from $\widehat{\mathbf{c}}$ (d) Perfectly
oriented powder where $\widehat{\mathbf{z}}\Vert \widehat{\mathbf{b}}$. For
the oriented powders the rf transmission is perpendicular to the $\widehat{%
\mathbf{c}}$ direction.}
\label{theoretcalnutation}
\end{figure}
\begin{figure}[tbp]
\begin{center}
\includegraphics[width=5cm]{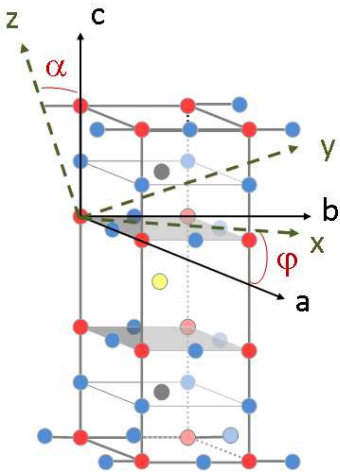}
\end{center}
\caption{(Color online) The unit cell of YBCO. An illustration of the
rotation of the EFG main principal axis from the crystallographic
directions, due to the missing oxygen atoms in the chains. The angle between $%
\widehat{\mathbf{z}}$ and $\widehat{\mathbf{c}}$ is $\protect\alpha $. The $%
\widehat{\mathbf{x}}$ direction is in the $a$-$b$ plane making an angle $%
\protect\varphi $ with $\widehat{\mathbf{a}}$}
\label{YBCOpicture}
\end{figure}
The Fourier transform part of this integral, over $t_{p}$, gives a delta
function at the angular frequency,
\begin{equation}
\omega _{p}(\theta ,\phi ,\eta )=\lambda (\theta ,\phi ,\eta )\omega _{1}.
\label{omegap}
\end{equation}%
The angular integrals required to complete the calculation of $I(\omega
_{p},t,\eta )$ demand numerical integration. Nevertheless, the peaks of this
function can be obtained by looking at the extremum of $\omega _{p}(\theta
,\phi ,\eta )$ with respect to $\theta $ and $\phi $. This gives three
singularities
\begin{equation}
\theta =0\qquad \omega _{I}=\frac{2\eta }{2\sqrt{3+\eta ^{2}}}\omega _{1}
\label{w1}
\end{equation}%
\begin{equation}
\theta =\pi /2\qquad \phi =0\qquad \omega _{II}=\frac{\eta +3}{2\sqrt{3+\eta
^{2}}}\omega _{1}  \label{w2}
\end{equation}%
\begin{equation}
\theta =\pi /2\qquad \phi =\pi /2\qquad \omega _{III}=\frac{\eta -3}{2\sqrt{%
3+\eta ^{2}}}\omega _{1}.  \label{w3}
\end{equation}%
In the special case of $\eta =0$ we find $\omega _{II}=\omega _{III}$ and $%
\omega _{I}=0$, and therefore there is only one sharp frequency at $%
0.866\omega _{1}$. As $\eta $ grows the three frequencies separate into the
three singularities and the pattern broadens. From these singularities $%
\eta $ can be extracted by $\eta =(\omega _{III}-\omega _{II})/(\omega
_{III}+\omega _{II})$. Figure~\ref{theoretcalnutation}(a) shows the powder
patterns for different values of $\eta $ for randomly oriented powder.

In the case of an oriented powder, the crystallographic $\widehat{\mathbf{c}}
$ direction and the EFG $\widehat{\mathbf{z}}$ direction are parallel but
the crystallographic $\widehat{\mathbf{a}}$ (and $\widehat{\mathbf{b}}$)
directions are random with respect to the EFG\ $\widehat{\mathbf{x}}$ (and $%
\widehat{\mathbf{y}}$) directions. The rf transmission is done perpendicular
to $\widehat{\mathbf{c}}$ (and $\widehat{\mathbf{z}}$). Figure~\ref%
{theoretcalnutation}(b) shows the nutation line shapes in this case. There
is no big difference between powder and oriented powder for $\eta =0$ but
as $\eta $ grows the difference between the two cases becomes clear.

In general, however, $\widehat{\mathbf{z}}$ does not have to be exactly
parallel to the $\widehat{\mathbf{c}}$ direction of the lattice. This
situation is demonstrated in Fig.~\ref{YBCOpicture}. The angle between $%
\widehat{\mathbf{z}}$ and $\widehat{\mathbf{c}}$ is $\alpha $. The $\widehat{%
\mathbf{x}}$ direction is assumed to be in the $a$-$b$ plane making an angle
$\varphi $ with $\widehat{\mathbf{a}}$. Figure~\ref{theoretcalnutation}(c)
shows the line shapes when $\alpha =20^{\circ }$ and $\varphi $ is
averaged. One can see that with the nutation spectroscopy it is very
difficult to distinguish between the case of perfect orientation (panel b)
and partial orientation (panel c). Finally, in Fig.~\ref{theoretcalnutation}%
(d) we present a case of perfectly oriented powder but with $\widehat{%
\mathbf{z}}$ parallel to $\widehat{\mathbf{b}}$. This case is fundamentally
different from previous cases especially when $\eta =1$.

For a spin-echo sequence, after a time $\tau $ from the first pulse there is
a refocusing pulse with duration $t_{r}$. This adds an additional factor of $%
\sin ^{2}[\lambda (\theta ,\phi ,\eta )\omega _{1}t_{r}]$ to the signal. In
an NQR experiment, since the spin rotation frequency depends on the
orientation of the lattice with respect to the coil, the second pulse cannot
perfectly refocus all the magnetization. Harbison \emph{et al}.\cite{Harbison2} showed that the additional factor does not change the nutation
frequencies $\omega _{I,II,III}$; however, it does change the relative
intensities of these frequencies with respect to each other.

The main advantage of the nutation method is that it is relatively simple to
execute; it can be carried out on a simple NQR spectrometer without a static
magnetic field or additional modifications. This method can also be
implemented for powders. It allows the determination of $\eta $ at every
point of the NQR spectrum (unlike NMR, where $\eta $ can be determined only
from the entire spectrum with no local resolution).

\subsection{Angle dependent NQR}
\begin{figure}[th]
\begin{center}
\includegraphics[width=8cm]{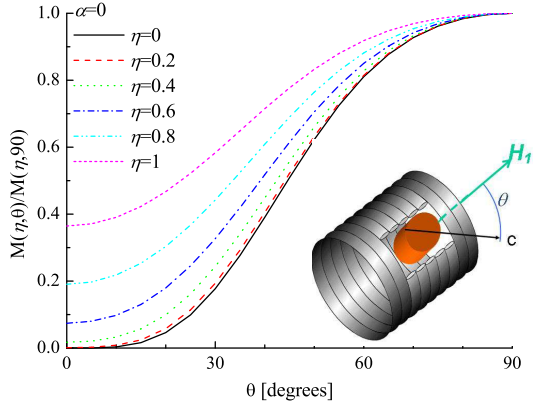}
\end{center}
\caption{(Color online) Theoretical echo intensity curves for various values
of $\protect\eta $ calculated from Eq.(\protect\ref{M}). $\protect\theta $
is the polar angle between the rf field and the $\widehat{\mathbf{c}}$
direction of the lattice, in an oriented powder. In this case the $\widehat{%
\mathbf{z}}$ direction of the EFG is assumed to be parallel to $\widehat{%
\mathbf{c}}$. Inset: basic angle-dependent NQR configuration. A sample with
a preferred direction is inserted into the coil. The angle $\protect\theta $
between them can be varied with a motor.}
\label{ADNQRdifferenteta}
\end{figure}

The angle-dependent NQR technique on an oriented powder was originally
developed by Levy and Keren.\cite{shahar} In this technique the signal
intensity for a given frequency is measured as a function of $\theta $, the
angle between the direction of the rf field and the crystallographic $%
\widehat{\mathbf{c}}$ direction. This technique is sufficient to determine $%
\eta $ when the EFG $\widehat{\mathbf{z}}$ direction is in the
crystallographic $\widehat{\mathbf{c}}$ direction. As we mentioned before
this does not have to be the case. In the ADNQR experiment the sample is
rotated with respect to the symmetry axis of the coil (see inset of Fig.~\ref%
{ADNQRdifferenteta}), and the echo intensity is registered for each angle.

From Eqs. (\ref{HQ}), (\ref{Hamrf}), and (\ref{Hrf}) one can see that for $%
\eta =0$, $\mathcal{H}_{Q}$, and $\mathcal{H}_{rf}$ commute when $\theta =0$.
In this case there will be no spin transitions and no signal. For $\eta >0$,
$\mathcal{H}_{Q}$, and $\mathcal{H}_{rf}$ do not commute even for $\theta =0$%
. In this case we expect a signal even when $\mathbf{H}_{1}$ is in the $\hat{%
z}$ direction. In general, Levy and Keren \cite{shahar} showed that for a $%
\pi /2-\tau -\pi $ pulse sequence, when $\mathbf{H}_{1}$ is given by Eq.(~\ref%
{Hrf}), the magnetization in the coil at the time of the echo is given by
\begin{equation}
M(\theta ,\phi ,\eta )=\frac{\lambda (\theta ,\phi ,\eta )\hbar \omega _{Q}}{%
2K_{B}T}\sin ^{3}{(\lambda (\theta ,\phi ,\eta )\omega _{1}t_{\pi /2})},
\end{equation}%
where $t_{\pi /2}$ is the duration of the $\pi /2$ pulse, and $\lambda
(\theta ,\phi ,\eta )$ is the same as in Eq.~(\ref{lambda}).

In the case of an oriented powder, again with $\widehat{\mathbf{z}}$ and $%
\mathbf{\hat{c}}$ parallel, and the $\mathbf{\hat{a}}$ (and $\mathbf{\hat{b}}
$) random, $M$ is obtained by averaging over $\phi $, namely,
\begin{equation}
M(\eta ,\theta )=\frac{1}{2\pi }\int_{0}^{2\pi }M(\theta ,\phi ,\eta ){d\phi
}.  \label{M}
\end{equation}%
Theoretical echo intensity curves as a function of $\theta $ for various
values of $\eta $ are presented in Fig.~\ref{ADNQRdifferenteta}. A fit of
experimental data to these theoretical curves can give the value of $\eta $.
\begin{figure}[tbp]
\begin{center}
\includegraphics[width=9cm]{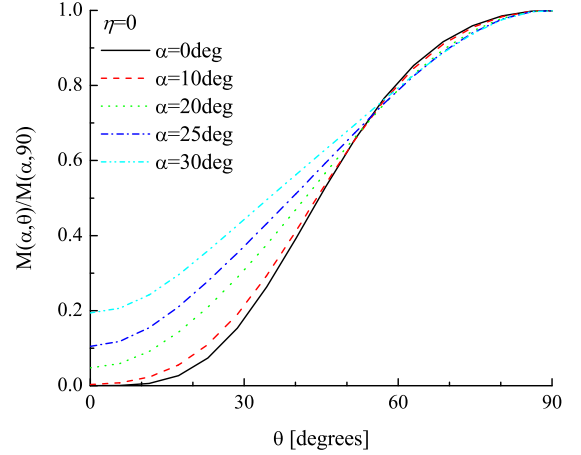}
\end{center}
\caption{(Color online) Theoretical echo intensity curves for $\protect\eta %
=0$ and various values of $\protect\alpha $ calculated from Eq.~(\protect\ref%
{Malpha}). $\protect\theta $ is the polar angle between the rf field and the
$c$ direction of the lattice, in an oriented powder.}
\label{ADNQRdifferentalpha}
\end{figure}

Next we discuss the case where $\widehat{\mathbf{z}}$ has an angle $\alpha $
with $\widehat{\mathbf{c}}$ as in Fig.~\ref{YBCOpicture}. In this case $%
\mathcal{H}_{Q}$ and $\mathcal{H}_{rf}$ again do not commute even for $%
\theta =0$ and we expect a signal even when $\mathbf{H}_{1}$ is in the $\hat{%
z}$ direction, for all values of $\eta $. To evaluate the signal intensity
we can take the unit vector pointing in the direction of $\mathbf{H}_{1}$ in
the crystal coordinate system to be $\widehat{\mathbf{r}}=\widehat{\mathbf{x}%
}\sin \theta +\widehat{\mathbf{z}}\cos \theta $ and express it as $\widehat{%
\mathbf{r}}^{,}$ in the EFG coordinate system using the Euler angles \cite%
{Euler}. If we assume, again for simplicity, that $\widehat{\mathbf{x}}$ is
in the $a$-$b$ plane with an angle $\varphi $ from $\widehat{\mathbf{a}}$ as
in Fig.~\ref{YBCOpicture} then,
\begin{equation}
\mathbf{{\mathord{\buildrel{\lower3pt\hbox{$\scriptscriptstyle\frown$}}\over
r}^{\prime }}}(\alpha ,\varphi ,\theta )=\left[ {\matrix{ 1 & 0 & 0 \cr 0 &
{\cos \alpha } & {\sin \alpha } \cr 0 & { - \sin \alpha } & {\cos \alpha }
\cr }}\right] \left[ {\matrix{ {\cos \varphi } & {\sin \varphi } & 0 \cr { -
\sin \varphi } & {\cos \varphi } & 0 \cr 0 & 0 & 1 \cr }}\right] \mathbf{{%
\mathord{\buildrel{\lower3pt\hbox{$\scriptscriptstyle\frown$}}\over r}^{\prime }}}%
(\theta )
\end{equation}%
and
\begin{eqnarray}
\lambda (\alpha ,\varphi ,\theta ,\eta )=
\end{eqnarray}
\begin{eqnarray}
\sqrt{\widehat{r}_{x}^{,2}(\alpha
,\varphi ,\theta )a_{x}^{2}(\eta )+\widehat{r}_{y}^{,2}(\alpha ,\varphi
,\theta )a_{y}^{2}(\eta )+\widehat{r}_{z}^{,2}(\alpha ,\varphi ,\theta
)a_{z}^{2}(\eta )}.\nonumber
\end{eqnarray}
In the case of $\eta =0$,
\begin{eqnarray}
\lambda (\alpha ,\varphi ,\theta ,\eta =0)=
\end{eqnarray}
\begin{eqnarray}
{\frac{\sqrt{3}}{2}}\sqrt{\cos
^{2}\varphi \sin ^{2}\theta +(\sin \alpha \cos \theta -\cos \alpha \sin
\varphi \sin \theta )^{2}}\nonumber
\end{eqnarray}%
and the magnetization in the coil as a function of the rotation angle $%
\theta $ is now
\begin{eqnarray}
M(\alpha ,\theta ,\eta =0)=\frac{\hbar \omega _{Q}}{2k_{B}T}\label{Malpha}
\end{eqnarray}
\begin{eqnarray}
\times\frac{1}{2\pi }%
\int\limits_{0}^{2\pi }{\lambda (\alpha ,\varphi ,\theta ,\eta =0)\sin ^{3}%
\left[ {\lambda (\alpha ,\varphi ,\theta ,\eta =0)\omega }_{1}{t_{\pi /2}}%
\right] d\varphi .} \nonumber
\end{eqnarray}%
Figure~\ref{ADNQRdifferentalpha} shows the magnetization in the coil according
to Eq.~(\ref{Malpha}) for different values of alpha.

\section{The experimental setup}

\label{Setup}

The NQR spectra were obtained using a spectrometer with a home-made
automated frequency sweep. The measurements were performed in a coil tunable
from 25 to 33MHz. The data were obtained by applying a spin-echo sequence.
Both for the nutation and the ADNQR techniques it is of great importance for
the rf field to be homogeneous. For the nutation experiment we measured
samples with a small volume inside a long cylindrical coil. The coil's wire
goes through a current monitor. This current monitor allows us to perform
measurements at different frequencies with the same current through the
coil and therefore the same $H_{1}$. For the ADNQR we used a spherical coil
that gives a more homogenous magnetic field with a better filling factor.
The experiment is fully automated, the sample holder is connected to a motor
which rotates the sample and can be controlled from the computer.

The measurements were done on YBCO$_{y}$ oriented powders, with different
doping levels. The data were taken at 100~K, which is above $T_{c}$ for all
samples. Previous Cu NQR measurements on YBCO (see Ref.\cite{NQR YBCO}) were performed on samples containing both Cu isotopes, Cu$%
^{63} $ and Cu$^{65}$, so the frequency lines consisted of doublets of Cu$%
^{63}$-Cu$^{65}$. In this work, for a clearer understanding of the NQR
signals, the YBCO samples were all made of enriched copper, meaning that
these samples contained only the Cu$^{63}$ isotope. This allows us to
distinguish between the different contributions to the NQR line from
different local environments.

\section{Results}

\label{Results}

\subsection{Cu NQR lines of enriched YBCO samples}

\label{sectionNQR}

\begin{figure}[tbp]
\begin{center}
\includegraphics[width=9cm]{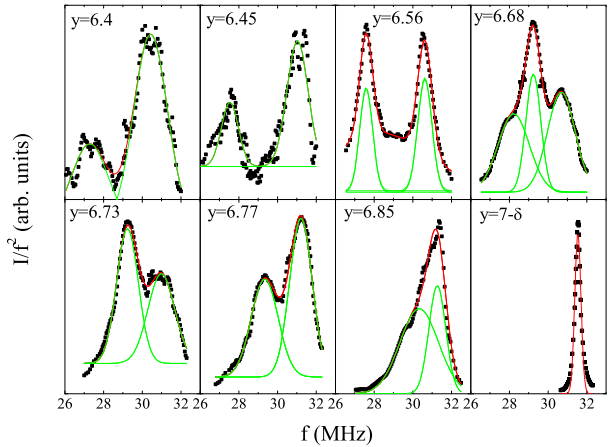}
\end{center}
\caption{(Color online) NQR frequency sweep on YBCO$_y$. The solid lines are
Gaussian fits performed in order to determine the resonance frequencies.}
\label{freqsweep}
\end{figure}

\begin{figure}[tbp]
\begin{center}
\includegraphics[width=8cm]{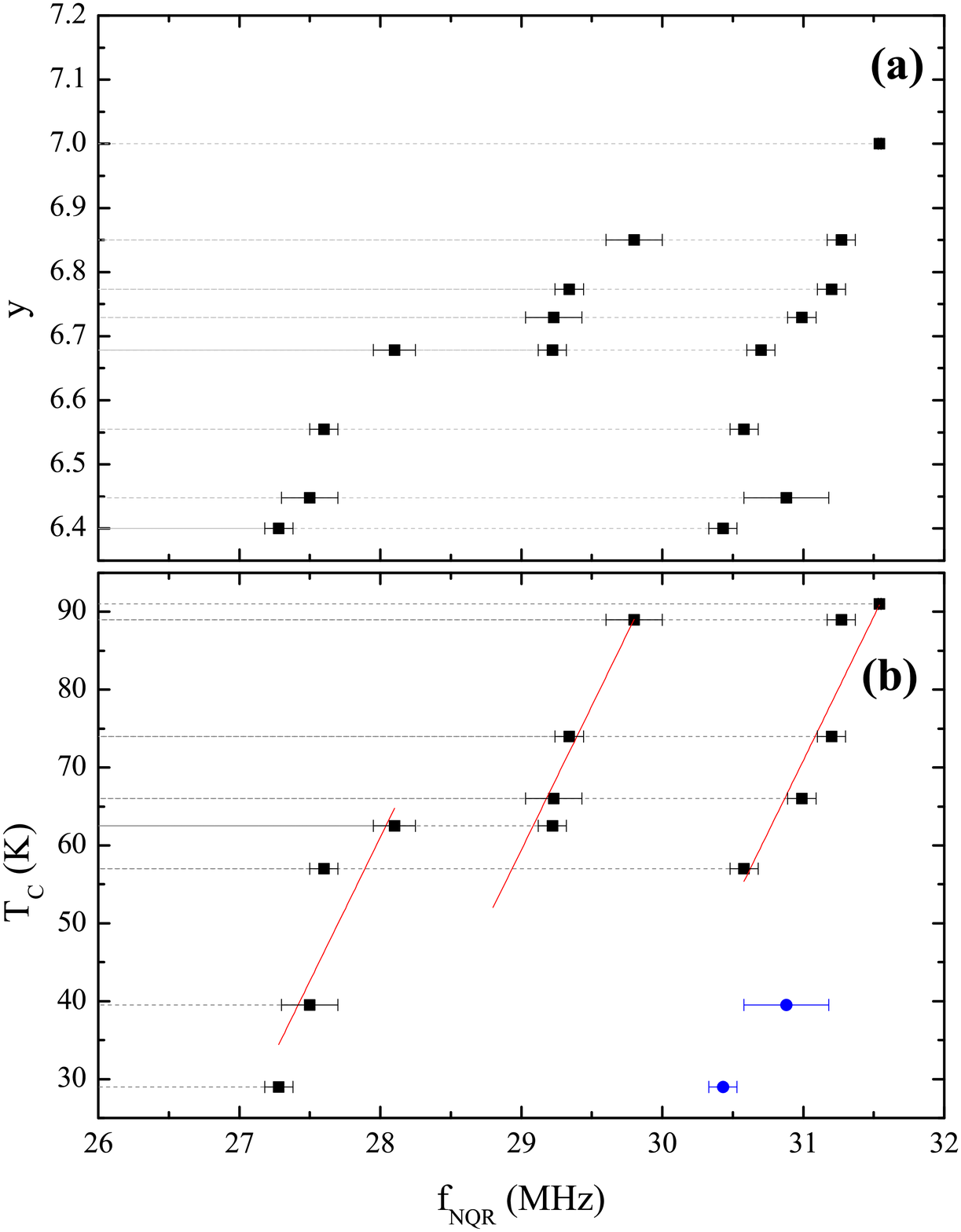}
\end{center}
\caption{(Color online) (a) The oxygen doping for different YBCO samples as
a function of the NQR frequencies. (b) The superconducting transition
temperature of these samples as a function of the NQR frequencies. The blue
circular symbols are resonance frequencies coming from the Cu(1) site. The
red lines are a fit to three parallel lines, for the three different Cu(2)
sites.}
\label{NQrfreq}
\end{figure}

The frequency sweep lines of YBCO$_{y}$ samples with different doping at $%
100 $~K are presented in Fig.~\ref{freqsweep}. The spectrum is normalized by
$f^{2}$ in order to correct for population difference and the induced signal
in the coil. For the fully doped sample only one resonance is observed. Between $y=6.85$
and $6.73$ two resonances are observed. At $y=6.68$ there are clearly three
resonances. The middle one disappears upon further oxygen reduction and
below $y=6.45$ again only two resonances are seen. For each sample, the
resonance frequencies were extracted from the spectrum. In Fig.~\ref{NQrfreq}%
(a) we plot the oxygen doping as a function of these frequencies. From this
plot it is clear that there are three different Cu sites.

Earlier NQR measurements done on YBCO$_y$ with both Cu isotopes present, found similar resonance frequencies for
similar $y$ values.\cite{NQR YBCO, Pennington} We therefore assign the different resonance frequencies of each sample to different
ionic environment, based on these earlier work. For high doping level ($y>6.5$) all signals
come from the plane Cu(2). There are three different types of environment
that affect the Cu(2) resonance frequency. They were classified in terms of
the number of oxygen surrounding the chain copper Cu(1) neighboring the
detected Cu(2). Cu(1)$_{4}$ stands for a Cu(2) next to a full chain as in
YBCO$_{7}$; in this case the frequency is the highest. Cu(1)$_{3}$ means
Cu(2) whose neighboring Cu(1) is missing one oxygen; the middle frequency
belongs to this Cu(2). Finally, Cu(1)$_{2}$ is when the neighboring chain to
the Cu(2) is empty; the lowest frequency belongs to this case. The three
possible environments of the Cu(1) are shown schematically in Fig.~\ref%
{Cusites}.

It is clear from Fig.~\ref{freqsweep} that as the doping decreases, the amplitude of the lower frequencies resonances increases at the expense of the high-frequency resonance. This makes sense with our line assignment, as at lower doping there are less Cu(2)$_{4}$ sites and more Cu(2)$_{3}$ and Cu(2)$_{2}$ sites. In addition, at oxygen doping close to 6.5 (i.e., 6.56 and 6.45) the signal from Cu(2)$_{3}$ is very hard to detect. This is due to the formation of the ortho II phase at these dopings.\cite{ortho II}

In Fig.~\ref{NQrfreq}(a) we depict the oxygen level $y$ vs. $f_{NQR}$.
Clearly, the more holes are introduced into the planes the more shifted is
the frequency. In Fig.~\ref{NQrfreq}(b) we plot the superconducting
transition temperature $T_{C}$ as a function of $f_{NQR}$. For each site,
the lower $T_{C}$ the more shifted are the peaks toward low frequencies. It
seems that the rate of change of $T_{C}$ with $f_{NQR}$; $dT_{C}/df_{NQR}$
is identical for all sites as demonstrated by the solid lines. Assuming that
the NQR resonance frequency is proportional to the number of charges (which
holds in the case where $\eta $ is constant), this implies that $T_{C}$ is
proportional to the number of free charges in the plane. Since $T_{C}$ is
also proportional to the superfluid density $n_{s}$,\cite{MuSRMe} this means
that the rate of conversion of holes to superconducting holes is constant.

\begin{figure}[tbp]
\begin{center}
\includegraphics[width=8cm]{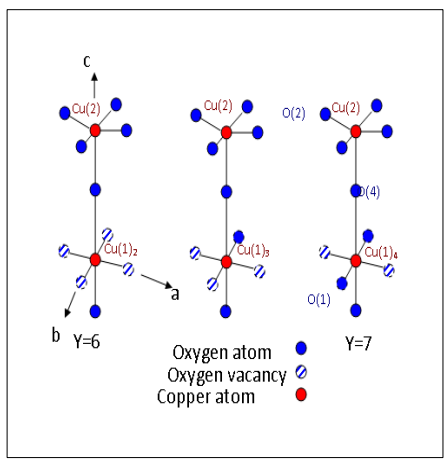}
\end{center}
\caption{(Color online) Schematic illustration of the Cu site in YBCO with
locally different oxygen coordinations.}
\label{Cusites}
\end{figure}

\subsection{Nutation spectroscopy}

\begin{figure}[tbp]
\begin{center}
\includegraphics[width=9cm]{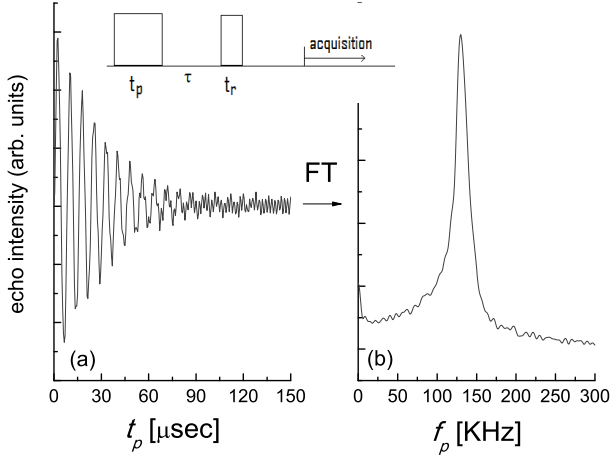}
\end{center}
\caption{Inset: the pulse sequence of a nutation experiment. (a) An example
of raw data of a nutation experiment for $y$=7-$\delta$ at 31.5MHz. The refocusing
pulse $t_{r}$ was 4.4$\protect\mu $s and $\protect\tau $ was 32$\protect%
\mu $s. (b) The data after Fourier transform.}
\label{rawnutation}
\end{figure}

In the nutation experiment we used a pulse sequence shown in the inset of
Fig.~\ref{rawnutation}. The measurement is done by applying this sequence at
the resonance frequency of each site, with varying excitation pulse length $%
t_{p}$. The entire echo is integrated and recorded as a function of $t_{p}$. An example for raw data for such experiment on a sample with $y$=7-$\delta$ is
shown in Fig.~\ref{rawnutation}(a). The Fourier transform over $t_{p}$ of
this data is presented in Fig.~\ref{rawnutation}(b).

\begin{figure}[tbp]
\begin{center}
\includegraphics[width=9cm]{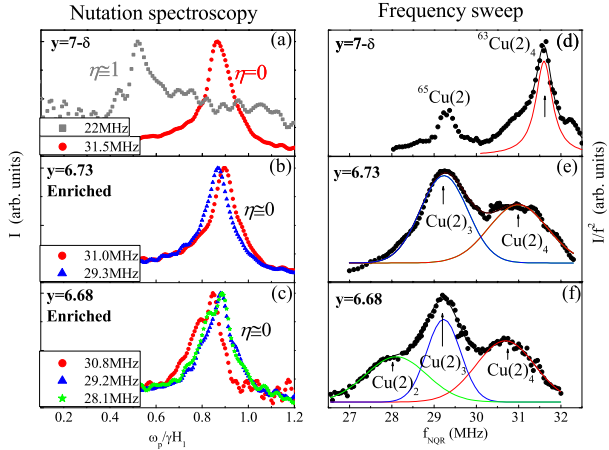}
\end{center}
\caption{(Color online) (a) Nutation spectra for YBCO$_{7-\delta}$ with natural
abundance of $^{65}$Cu and $^{63}$Cu. [(b) and (c)] Nutation spectra for YBCO$%
_{6.73}$ and YBCO$_{6.68}$ enriched with $^{63}$Cu. [(d)-(f)] Cu NQR
line shape for these three samples, extracted from Fig.~\protect\ref%
{freqsweep}. The arrows show the frequencies where nutation spectroscopy is
applied.}
\label{Nutationresults}
\end{figure}

For the Cu(2) of fully doped YBCO it was established \cite{Pennington, Shimizu} that the principal component of the EFG is in the $\widehat{%
\mathbf{c}}$ direction of the lattice and that $\eta =0$. Looking at Eqs.~(%
\ref{omegap})-(\ref{w3}), in the case of $\eta =0$ there is only one
nutation frequency and the ratio $\omega _{p}/\omega _{1}$ is 0.866.
Therefore, from the nutation frequency of the Cu(2) resonance of YBCO$_{7-\delta}$
we can extract $\omega _{1}$. Since we worked with a constant rf field $%
H_{1} $, we normalized the frequency axis for all samples by $\omega _{1}$.
We then apply this technique to three YBCO$_{y}$ samples. The results are
shown on the left panels of Fig.~\ref{Nutationresults}. For each sample the
measurements were done at the frequencies marked with arrows on the NQR
spectrum, extracted for clarity from Fig.~\ref{freqsweep}, and shown in the
right panels.

Figure~\ref{Nutationresults}(a) shows the nutation spectrum for YBCO$_{7-\delta}$,
measured in both the Cu(2) (31.5MHz) and the Cu(1) (22MHz) resonance
frequencies. After normalizing the frequency axis by $\omega _{1}$, as
explained above, the Cu(2) has a sharp nutation frequency with $\omega
_{p}/\omega _{1}=0.866$. However, Cu(1) has a much broader spectrum with $%
\omega _{pIII}/\omega _{1}=0.52$ and $\omega _{pII}$ which is difficult to
determine. The result for the Cu(2) is consistent with $\eta =0$ [see
theoretical nutation spectra in Fig.~\ref{theoretcalnutation}(b)]. The
result for Cu(1) is consistent with $\eta =0.95\pm 0.05$, and $\widehat{%
\mathbf{z}}$ pointing in the $\widehat{\mathbf{b}}$ direction, as shown in
Fig.~\ref{theoretcalnutation}(d). These results are in agreement with NMR
results on YBCO$_{7-\delta}$ (Refs.~\cite{Pennington, Shimizu}) that measured $\eta
\simeq 0$ for Cu(2), and $\eta \simeq 1$ for Cu(1). Similar nutation
experiments on powder YBCO$_{7-\delta}$ at room temperature were done by Vega.\cite{VegaNutation} His results are $\eta =0$ for the Cu(2) and $\eta =0.8$ for
the Cu(1).

Figures.~\ref{Nutationresults}(b) and \ref{Nutationresults}(c) show the nutation spectroscopy
results for YBCO$_{y}$ with lower doping levels. For these samples we
measured only at the resonance frequencies of the Cu(2). The sample with $%
y=6.73$ has two resonance lines for two different Cu(2) ionic environments,
and the sample with $y=6.68$ has three Cu(2) environments. The nutation
experiment shows that for these samples, for all three different types of
Cu(2) ionic environments, the nutation spectrum is not different from that
of the $y$=7-$\delta$ case, therefore $\eta \simeq 0$. This conclusion was verified
by using several different $H_{1}$ and refocusing pulse time $t_{r}$.
However, as mentioned in section \ref{sectionnutationtheory}, the nutation
experiment is not sensitive enough to small differences in the orientation
of the EFG [see Figs.~\ref{theoretcalnutation}(b) and \ref{theoretcalnutation}(c)].

\subsection{ADNQR}
\begin{figure}[tbp]
\begin{center}
\includegraphics[width=9cm]{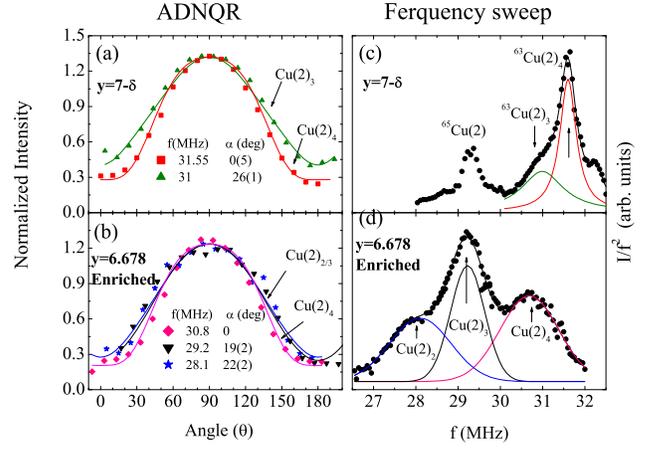}
\end{center}
\par
\caption{(Color online) Right panels: Cu NQR line shape for YBCO$_{7-\delta}$ with
natural abundance of $^{65}$Cu and $^{63}$Cu, and for YBCO$_{6.68}$ enriched
with $^{63}$Cu. The arrows show the frequencies where ADNQR is applied. Left
panels: the echo intensity as a function of the angle $\protect\theta $
between the rf field and the $\widehat{\mathbf{c}}$ direction of the
lattice, for the two samples. The solid lines are fits to Eq.~(\protect\ref%
{Malpha}).}
\label{ADNQRresults}
\end{figure}

As mentioned before, in a fully doped YBCO $\widehat{\mathbf{z}}$ is parallel to the
$\mathbf{\hat{c}}$ axis of the lattice. Therefore the ADNQR technique can be
applied to this sample, using Eq.~(\ref{M}). The result indeed gives $\eta
=0 $ as was published in Ref. \cite{ADNQR2006} and in agreement with the
nutation experiment described above. For lower doping levels however, the
exact direction of $\widehat{\mathbf{z}}$ is unknown. In a previous
publication in Ref.~\cite{ADNQR2006} we also performed an ADNQR experiment
on a YBCO sample with doping of $y=6.68$. Due to lack of information we
assumed that $\widehat{\mathbf{z}}\Vert \mathbf{\hat{c}}$ at all doping. The
experimental results from Ref.~\cite{ADNQR2006} are shown by the dotted line
in Fig.~\ref{ADNQRresults}(b). Fitting these data to Eq.~(\ref{M}) gave a
high value of $\eta $ in contradiction to the nutation experiment findings
in Fig.~\ref{Nutationresults}(c) of the present work. The way to settle this
contradiction is by allowing $\widehat{\mathbf{z}}$ to be tilted with an
angle $\alpha $ from $\mathbf{\hat{c}}$ as in Fig.~\ref{YBCOpicture}. As
mentioned above, in this case the NQR signal does not disappear at $\theta
=0 $, even in the case of $\eta =0$, as result of $\widehat{\mathbf{z}}$ EFG
component in the $a-b$ plane.

Figures.~\ref{ADNQRresults}(a) and \ref{ADNQRresults}(b) show the ADNQR data for the two YBCO
samples with doping of $y$=7-$\delta$ and $6.68$. The solid lines are new fits,
this time to Eq.~(\ref{Malpha}), namely, $\eta =0$ and tilted principle
axis. The fit allows for a finite base line for each sample to account for
some unknown amount of misalignment. This misalignment is a result of the
non-perfect orientation and some inhomogeneity of the induced field in the
coil. These samples' frequency spectrum, taken from Fig.~\ref{freqsweep},
are presented again in Figs.~\ref{ADNQRresults}(c) and ~\ref{ADNQRresults}(d) for clarity. The
arrows in this figure mark the frequencies where ADNQR was applied.

For the main peak of $y$=7-$\delta$ the best fit is obtained with $\alpha =0$ and $%
\eta =0$, as expected. The same result was obtained at the high frequency of
the $y=6.68$ sample, which is also associated with Cu(2) in an environment
of full chains; Cu(1)$_{4}$. In contrast, for the other oxygen environments
the signal at $\theta \rightarrow 0$ is clearly above the background.
Keeping in mined our nutation spectroscopy results, showing that $\eta =0$,
this suggests that the value of $\alpha $ is larger than zero. The fit to
Eq.~(\ref{Malpha}) gives $\alpha \simeq 20^{\circ }\pm5^{\circ }$. This is the main
finding of this work.

\section{Summary and conclusions}

\label{Conclusions}

We used nutation spectroscopy and ADNQR to measure the quadrupole
interaction asymmetry parameter $\eta $ and tilting angle $\alpha $ of the
main component of the EFG principal axis $\widehat{\textbf{z}}$ from the
crystallographic $\widehat{\textbf{c}}$ direction. Calculations show that nutation
spectroscopy powder patterns for oriented powder are very sensitive to the
value of $\eta $, but they are not very sensitive to $\alpha $ for $\alpha $
up to $20^{\circ }$. In contrast, the ADNQR line shape is sensitive to both $%
\eta $ and $\alpha $. In other words, ADNQR is sensitive to any breaking of
the axial symmetry. However, it is difficult to determine both $\alpha $ and
$\eta $ from ADNQR alone. The combination of the two methods can give a very
good estimate on both parameters. We implement these techniques on the
in-plane Cu atom of the YBCO compound with different doping levels. We
conclude that for both YBCO$_{7-\delta}$ and YBCO$_{6.68}$, $\eta $ is
approximately zero for all Cu(2) resonance frequencies. However, the ADNQR
showed that for the underdoped sample there is a clear difference between
the Cu(2) neighboring a Cu(1) in a full chain as in a fully doped sample [Cu(1)$_{4}$%
] and the Cu(2) neighboring a Cu(1) in an empty or a half filled chain
[Cu(1)$_{2}$-Cu(1)$_{3}$]. This difference can result from a small change in
the EFG principal axis $\widehat{\mathbf{z}}$ with respect to the lattice
directions $\widehat{\mathbf{c}}$.

The motivation for these experiments was to measure possible charge
inhomogeneity or electronic phase separation in the YBCO compound. Both the
nutation spectroscopy and the ADNQR for the Cu(2) from a fully oxygenized
local environment Cu(1)$_{4}$, even for lower averaged doping, show a
homogeneous charge distribution in the plane. Our combined experiments also
imply that only for Cu(2) neighboring a Cu chain with missing oxygen Cu(1)$%
_{2}$ or Cu(1)$_{3}$ is the principle axis $\widehat{\mathbf{z}}$ of the EFG
tensor not along $\widehat{\mathbf{c}}$, namely, $\alpha \simeq 20^{\circ }\pm5^{\circ }$. Therefore, only from the point of view of a Cu(2) in YBCO$%
_{6.68}$ neighboring a Cu(1)$_{2}$ or Cu(1)$_{3}$, are the $\widehat{\mathbf{%
a}}$ and $\widehat{\mathbf{b}}$ crystallographic directions less identical
than in YBCO$_{7}.$ This means that any charge inhomogeneity in the plane is
correlated directly with the O dopant atoms, and therefore cannot be an intrinsic property of the CuO$_2$ planes alone. It
is either generated or pined by missing oxygen.

McElroy \emph{et al.} \cite{McElroy2005} came to a similar conclusion by
performing spectroscopic imaging scanning tunneling microscopy on Bi-2212
samples. They found strong correlation between the position of localized
resonance at -960meV identified with interstitial oxygen dopants and the
size of the local spectral gap. To understand these result, Nunner \emph{et
al.} \cite{Nunner} presented a theoretical model where the dopants modulate
the pair interaction locally on an atomic scale. They calculated the
correlation between the local density of states and the dopant modulated
pair interaction potential. They showed that this model agrees with
McElroy's experimental results on Bi-2212. A more recent theoretical work by
Mori \emph{et al.} \cite{Mori} identified two mechanisms by which the
position of the apical oxygen atoms can modulate the pairing interaction within
the CuO$_{2}$ planes.

Our result for the YBCO compound reinforces the surface experiments done on
Bi-2212. It shows that the correlation between the electronic spatial
variation in the plane and the dopant exists not only in Bi-2212 and it is
a property of the bulk and not only of the surface.

\section{Acknowledgments}

This work was funded by the Israeli Science Foundation.

\end{document}